# The CMS Magnet Test and Cosmic Challenge

Tim Christiansen for the CMS Collaboration

*Abstract*–The CMS[1] detector is under construction for imminent operation at the Large Hadron Collider at CERN near Geneva, Switzerland. The installation and commissioning is performed in a surface hall. Thereafter, all the main heavy elements of the detector, the disks and wheels, are lowered in the experimental cavern starting at the end of 2006. The superconducting magnet of CMS requires testing before lowering, providing a unique opportunity to operate all the sub-detectors and sub-systems together and to take data with cosmic muons as an important commissioning test. This is called the MTCC – Magnet Test and Cosmic Challenge. The participating systems include a 60º sector of the Muon System comprising gas detectors like the drift tubes (DTs), Cathode Strip Chambers (CSCs) and Resistive Plate Chambers (RPCs), both in the Barrel and Endcaps. The tracking system comprises elements of the Silicon-Strip Tracker, and parts of the Electromagnetic and Hadronic Calorimeter detected energy depositions of the traversing muons. In this article, a description of the operational experience and the lessons learnt are presented.

## I. INTRODUCTION

THE main components of CMS [1] are the tracking system [2], made of silicon pixel and strip layers, finely segmented electromagnetic [3] and hadron [4] calorimeters, a large solenoid magnet with a 4T field [5], and, outside the magnet coil and the calorimeters, the muon detectors [6] and iron absorbers also serving as a magnet yoke, as shown in Fig. 1.

A full-scale test and field-mapping of the state-of-the-art CMS 4 Tesla solenoid magnet system has always been envisaged prior to lowering the major elements of the experiment into the underground experiment cavern. The eventual evolution of the schedule of the experimental and service caverns underground forced CMS to complete even more of the detector on the surface. Thus arose the concept of combining cosmic ray testing of several sub-detectors with the magnet test, which developed into the "Cosmic Challenge", a simultaneous system test of all parts of CMS.

The original objectives of the Magnet Test and Cosmic Challenge (MTCC) were stated as follows:
1. Test and commissioning of the Magnet, including cooling, power supply and control system. Mapping the magnetic field.
2. Check of the yoke closure system and tolerances, the movement under magnetic field and the functionality of the muon alignment system (Endcap + Barrel + link to Tracker).
3. Check of the magnetic field tolerance of yoke mounted components.
4. Check of the installation and cabling of HCAL, ECAL and Tracker inside the coil.
5. Combined test of sub-detectors in a 20º slice of CMS with magnet, using as near possible final readout, data acquisition (DAQ), trigger, control, safety and auxiliary systems. Check of noise and inter-operability of the subsystems. The goal was to trigger and record cosmic muons and to construct intelligible events in many sub-detectors simultaneously in order to test and optimize the commissioning, synchronization and operation procedures for CMS.

The benchmark of success in this last objective was considered to be the recording, offline reconstruction and display of a cosmic ray in the four subsystems of CMS (Tracker, ECAL, HCAL and Muon Detector) with the magnet operating at 4T.

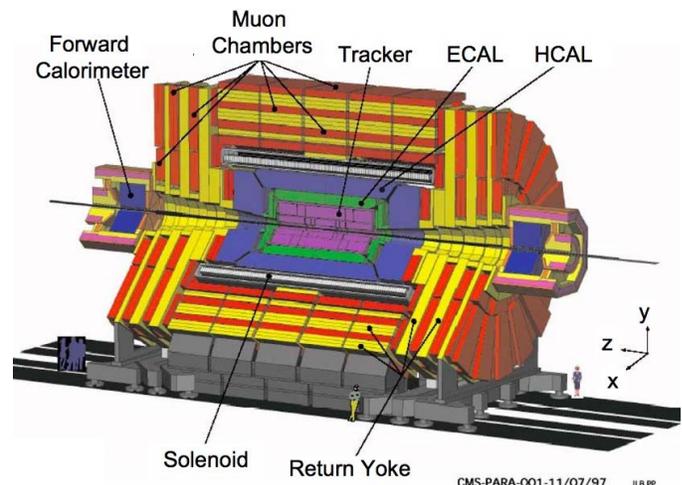

Fig. 1. Schematic view of the CMS detector [1].

The MTCC was split in two distinct phases of data taking: In Phase I, in parallel to the magnet commissioning, the Tracker and ECAL components inside the HCAL were included in global data taking, thus allowing the collection of cosmic muons passing all inner and outer sub-detectors. For the second phase, Phase II, the Field Mapper replaced the ECAL and Tracker detectors, while the HCAL and the muon detectors continued to record cosmic events in parallel to the mapping of the magnetic field.

Additional constraints of the cosmic data taking were not to introduce any significant delay to the assembly progress, to perform the "cosmic challenge" in the shadow of commissioning and field-mapping, to complement the trigger system (high rate) tests in the CMS electronics integration centre, and to use final systems as far as possible with no (or few) MTCC specific developments.

The original scope was expanded to include substantial offline as well as online systems objectives. Data transfer to some Tier-1 centers, online event display, "quasi-online" data-quality analysis, and fast offline data-checking at Fermilab were some highlighted targets of the MTCC Phase I, designed to offer a first hand taste of a 'CMS-like' running experience.

## II. MAGNET TEST AND DETECTOR PREPARATION

The muon Drift Tubes (DT) setup for the MTCC consisted of 14 stations distributed on four layers in three neighboring bottom sectors at the +z side of the Barrel detector. Each station provides measurements to build two φ-view segments and one θ-view segment. Thirty-six Cathode Strip Chambers (CSC) chambers forming a 60º slice corresponding to the same azimuth coverage of the DT system, complement the active muon detectors in the +z Endcap. Each CSC chamber can provide one three-dimensional track segment. Two Electromagnetic Calorimeter (ECAL) super-modules and 15 Hadron Calorimeter (HCAL) wedges were powered and read out during the MTCC. The MTCC setup is completed with 133 single and double-sided Silicon Strip Tracker modules. Up to six Tracker layers can be traversed by a straight track.

After the cabling of the detector elements, the final closing of the yoke proceeded. It was completed in July 2006, marking the beginning of the magnet test. The magnet power circuit was completed and the testing of the coil progressed in steps if increasing current: 5 kA, 7.5 kA, 10 kA, 12.5 kA, 15 kA, 17.5 kA, and then 19.12 kA to reach the nominal 4T field on 22 August 2006 (see Fig. 2). At each new field value, fast discharges were provoked to learn how to tame the dumping of energy inside the cold mass during such a discharge. At 12.5 kA a rupture disk was broken due to bad functioning of a safety valve. However, after adjustment of the cold box process to limit the quantity of helium in the phase separator circuits, the last fast discharges from the nominal current behaved as expected. All other parameters registered nominal values and the electrical insulation of the coil is good and all cooling circuits are vacuum tight. After running at 3.8T for 48 hours for the cosmic challenge (see Fig. 2), the coil was operated at 4 T for 2 hours on the 28 August, and the functional tests of the magnet have been declared completed with success.

After having extracted the ECAL and Tracker components, the Field Mapper was installed inside the HCAL volume and the detector was closed again in preparation for the MTCC Phase II. During the following three weeks in October/November 2006, the magnetic field inside the solenoid was mapped at various nominal field points, 0 T, 2 T, 3 T, 3.5 T, 3.8 T, and 4 T, with a precision of the order of $10^{-4}$.

## III. THE COSMIC CHALLENGE

Twenty five million cosmic triggered events were recorded with the principal sub-detectors active during Phase I, of which 15 million events have stable field $B \geq 3.8$ T, as can be seen in Fig. 3. The cosmic-muon trigger consisted of a combination of triggers from all muon systems, yielding a total rate between 100 and 300 Hz, depending on the trigger configuration. Data-taking efficiency reached over 90% for extended periods. Several thousand of these events have reconstructed tracks form clusters in three or more layers of the central Tracker. One of the first of such events recorded is shown in Fig. 4. The analysis of the whole data sample will provide useful understanding and calibration of the combined detector and software performance. Fig. 5 shows preliminary cosmic-muon distributions, reconstructed from signals in the innermost layer of the DT chambers at $B = 0$ T and $B = 3.5$ T, compared to a simulation for $B = 0$ T and $B = 4$ T, respectively.

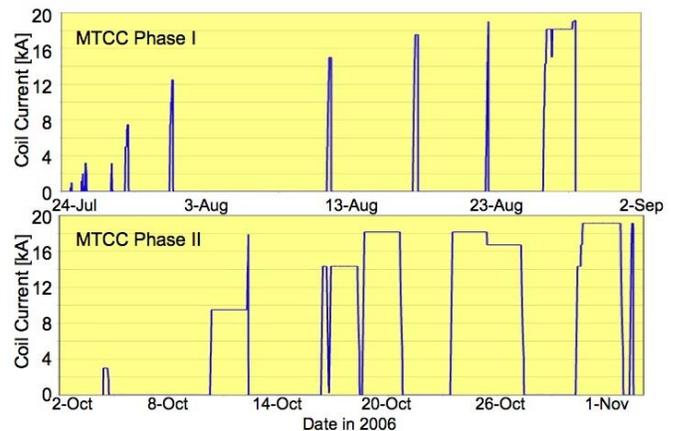

Fig. 2. Magnet current versus time during the two phases of the MTCC.

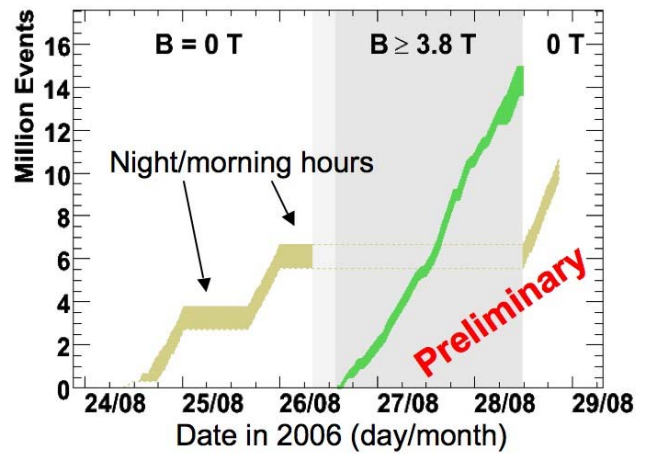

Fig. 3. Number of accumulated events in the first phase of the MTCC with the muon systems (Drift Tubes for the CMS Barrel and Cathode Strip Chambers for the Endcap), the hadron and electromagnetic calorimeters, and the central Tracker modules in synchronized global readout. The diagram shows the events for runs with and without magnetic field $B$.

With the first phase of the MTCC successfully completed, ECAL and Tracker elements had been removed to make way for the Field Mapper and corrections had also been made for field intolerant features found during MTCC Phase I in order to benefit from further testing and verification during the Phase II. The principle objectives of the continued data taking with the muon systems and the HCAL in Phase II were to study the detailed response of HCAL and the muon detectors to the magnetic field, and to commission newly available elements of the global trigger and the RPC readout system. In addition to the muon triggers from the DT, CSC and RPC systems, an additional HCAL trigger has been commissioned for the Phase II, based on the coincidence of signals from minimum-ionizing particles in the upper and lower part of the HCAL detector.

Additional tens of millions of events have been recorded for various magnetic fields and trigger configurations during Phase II, which are currently being analyzed.

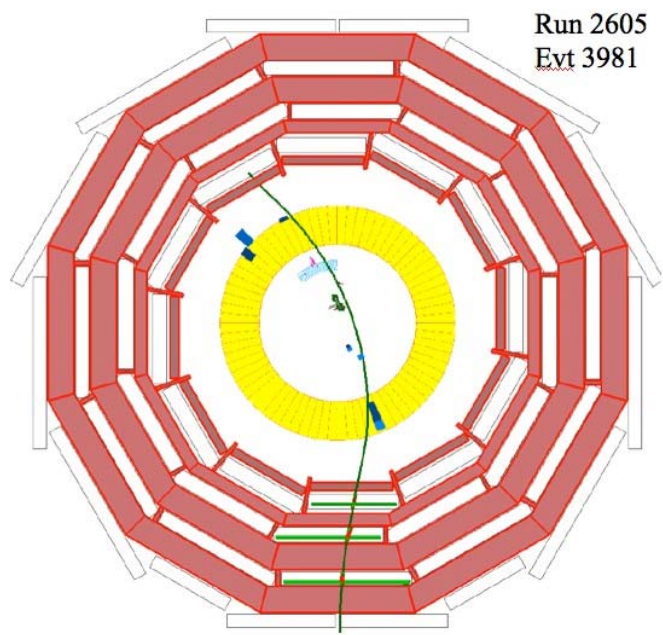

Fig. 4. Event display of a reconstructed cosmic track through all subsystem layers of the Barrel detector as seen from the Endcap. The green s-shaped curve indicates the result of a reconstruction of the cosmic track in the muon system. This event, which was recorded on 6 August 2006 with a central magnetic field of $B \approx 3.5$ T, was triggered by a coincidence of signals in multiple layers of the Drift Tube system (outer-most ring, bottom sector). In line with the resulting track, corresponding signals can be identified in the hadron calorimeter (blue bars on the yellow ring representing the Barrel HCAL), the electromagnetic calorimeter (see pink cluster in the active ECAL modules in the 11 o'clock position), and in multiple layers of the Silicon-Strip Tracker (green), located just above the center of the detector.

IV. SUMMARY AND CONCLUSIONS

After 15 years starting from early design, R&D, pre-industrialization, six years of construction and about one year of installation, CMS coil has been tested successfully. From cryogenic, electrical and mechanical tests, the coil fulfilled all specifications and seemed easy to operate, thus allowing to proceed with the installation of CMS detectors inside the experimental area as planned.

The main conclusions of the Magnet Test and Cosmic Challenge were that CMS could be opened and closed on the timescales intended, that the magnet worked stably and safely at the nominal field around 4T, that the sub-detectors coherently worked with the magnet and each other, that the sub-detectors could be integrated with the central systems like DAQ, trigger, detector safety and control systems, the central data-quality monitoring system etc., that precise three-dimensional maps of the solenoid field inside the HCAL were produced for various coil currents, that the commissioning strategy broadly worked, and that CMS started to work as a worldwide, unified team.

As a result of the diligent work by hundreds of people over many years, the MTCC concluded on time with all objectives met.

While the $+z$ end of the CMS detector is lowered into the detector cavern first, almost all the MTCC infrastructure in surface building will remain in place until the middle of 2007, to act as a commissioning platform for the $-z$ end elements and as a further test-bed for integration with trigger and DAQ.

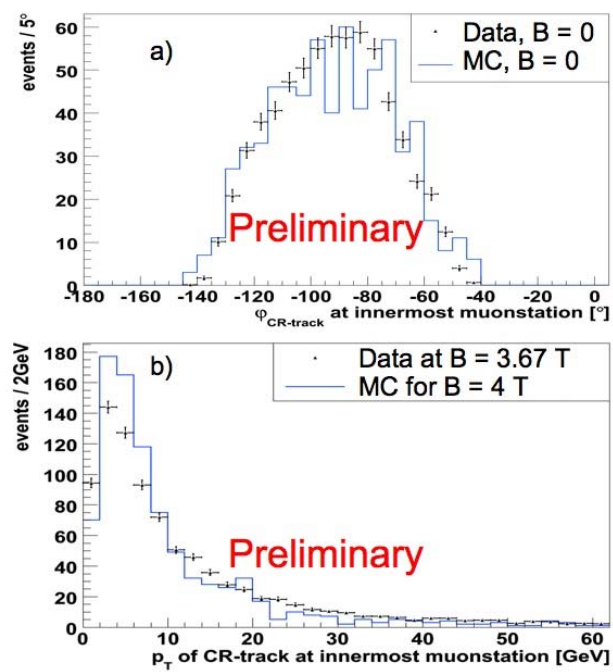

Fig. 5. Preliminary cosmic muon spectra as reconstructed in the innermost layer of Drift Tube chambers as compared to a Monte Carlo simulation (MC): a) azimuth angle $\varphi$ for magnet field $B = 0$; b) reconstructed "transverse momentum" $P_T$ for magnetic field of 3.67 T compared to MC with 4 T nominal field.